\documentstyle[preprint,aps,epsfig]{revtex}
\begin{document}
\draft
\preprint{\vbox{\hbox{CERN-TH/97-263}
                \hbox{hep-th/9710070}}}
\title{Hamilton-Jacobi Approach to Pre-Big Bang Cosmology
at long-wavelengths}
\author{\footnote{This work is dedicated to K. At{\i}lan.}K. Saygili }
\address{Theory Division, CERN, CH-1211 Geneva 23, Switzerland \\ 
\vspace{.2in}
\footnote{After 10 Sep. 1997.\\ \hspace*{.1in} e-mail: saygili@boun.edu.tr}
Bo\~gazi{\c c}i University, Department of Physics\\
80815, Bebek, Istanbul, Turkey}
\maketitle

\begin{abstract}
 
    We apply the long-wavelength approximation to the
low-energy effective string action in the context of Hamilton-Jacobi
theory. The Hamilton-Jacobi equation for the effective string action 
is explicitly invariant under scale factor duality.
We present the leading order, general solution of the Hamilton-Jacobi 
equation. The Hamilton-Jacobi approach yields a solution consistent with the 
Lagrange formalism. The momentum constraints take an 
elegant, simple form. Furthermore, this general
solution reduces to the quasi-isotropic one, if the evolution of the 
gravitational radiation is neglected. Duality transformation for the general 
solution is written as a coordinate transformation \enlargethispage*{1000pt}
in an abstract field space.

\end{abstract}

\section{Introduction}

     In this paper, we apply the long wavelength approximation to the
low-energy effective string action in the context of Hamilton-Jacobi
theory.

      The correct theory of quantum gravity is generally believed to be
string theory. The low energy effective action of string theory
has the form of a Brans-Dicke (BD) action with parameter
$\omega_{BD}=-1$. Duality symmetries play an important role in string theory. 
In the Pre-Big Bang scenario \cite{art1} scale factor duality associates
inflationary solutions to non-inflationary ones, \cite{art2}. By this
property it has become an alternative to standard inflationary
cosmology.

     Long-wavelength gravity (gradient expansion) has proved to be a
fruithful method for studying (slightly) inhomogeneous
fields. It is a significant improvement over those of homogeneous
minisuperspace. The main idea of this scheme is that when the scale of spatial
variations of the fields are larger than the Hubble radius, one can solve
the equations neglecting the second-order spatial gradients. It was
first introduced by E. M. Lifschitz and I. M. Khalatnikov \cite{art3}.
Later, Tomita developed this approximation as the Anti-Newtonian scheme
\cite{art4}. For General Relativity, it was formulated either directly
in terms of lagrange equations \cite{art5}, or in the framework of 
Hamilton-Jacobi equation \cite{art7,art6}. It was also studied for the 
BD theory in the Hamilton-Jacobi framework \cite{art8}.

Recently, an inhomogeneous version of Pre-Big Bang cosmology has been
investigated using the lagrange equations \cite{art9,art11}. In this
scenario, the universe, with very perturbative (i.e. weak coupling and
very small curvature) but otherwise arbitrary initial conditions is
followed towards the Big Bang singularity in the future. Quasi-homogeneous
regions, which exhibit Pre-Big Bang behaviour, eventually fill almost all
of space, and within these regions the Universe appears homogeneous, flat,
and isotropic \cite{art9}. 

     The purpose of the present paper is two-fold. First we solve the
long-wavelength problem for low-energy string cosmology in the framework
of Hamilton-Jacobi (HJ) equation. We also show that the HJ equation is
invariant under scale factor duality (SFD) transformation. We work
directly in the physical string frame. The HJ approach has importance for
quantum cosmology since it is the lowest order equation in the WKB
approximation of Wheeler-De Witt equation. Secondly, following Salopek
\cite{art7}, we write the momentum constraints in a simple, elegant form. 

      In section II, we write the action in the hamiltonian form and we
derive the equations for the fields and for their conjugate momenta. The
action also gives rise to Hamiltonian and the momentum constraints. For
completeness, we also give a brief summary of the canonical
transformations via the generating functional technics. In section III, we
write the Hamilton-Jacobi equation, and we show that it is invariant under
SFD transformation.  The solutions represent a universe evolving towards a
Big Bang singularity in the future. Section IV includes the most general
solution near a singularity. In this section, the generating functional is
taken as a function of both the dilaton and the metric. This dependence is
chosen in such a way that will effectively decompose the gravitational
momentum tensor into a trace contribution and a traceless part.
Furthermore, with the help of this choice, the Hamilton-Jacobi equation
reduces to that of massless scalar fields.  We write the momentum
constraint in a simple form. The momentum constraints state that the
generating functional is invariant under spatial coordinate
transformations. We write them in terms of the new canonical variables.
Then in section V, we present a quasi-isotropic solution of Pre-Big Bang
cosmology. The general solution reduces to the quasi-isotropic one, if the
evolution of the gravitational radiation is neglected.  In section VI, we
briefly discuss duality transformation for the general solution. We
represent the evolution of the universe in a space of fields where the
duality transformation can be written as the transformation of an angle in
a suitable plane. 

\section{Hamilton Formalism and the Canonical Transformations}

     The low-energy effective string action, in the string frame, is 

\begin{equation}\label{eq:1} 
\Gamma=\int e^{-\phi}(R+
g^{\mu\nu}\partial_{\mu}\phi
\partial_{\nu}{\phi}) \sqrt{-g}d^{4}x ~~~,
\end{equation}

\noindent where $R$ is the scalar curvature, $\phi$ is the dilaton. We set
to zero the antisymmetric tensor field $B_{\mu\nu}$. The HJ equation for
this action can be obtained using the ADM formalism in which the
space-time is foliated by space-like hypersurfaces. In the ADM formalism
the metric is parametrized as

\begin{equation}\label{eq:2}
ds^{2} =
-N^{2}(t)dt^{2}+\gamma_{ij}(dx^{i}+N^{i}dt)(dx^{j}+N^{j}dt) ~~~,
\end{equation}

\noindent  where $N$ and $N^{i}$ are the lapse and shift functions
respectively, and $\gamma_{ij}$ is the $3-$metric. We work in the 
synchronous gauge $N=1$, $N^{i}=0$.

    The action written in hamiltonian form becomes,

\begin{equation}\label{eq:3}
\Gamma=\int(\pi^{ij}\dot\gamma_{ij}+\pi^{\phi}\dot\phi-N{\mathcal{H}}
-N^{i}{\mathcal{H}}_{i})d^{4}x ~~~,
\end{equation}

\noindent  where $N$, and $N^{i}$ act as lagrange multipliers. 
Their variation gives rise to the hamiltonian constraint

\begin{eqnarray}\label{eq:4}
{\mathcal{H}} & = &
\gamma^{-1/2}e^{\phi}[\pi^{ij}\pi^{kl}\gamma_{ik}\gamma_{jl}
+\frac{1}{2}(\pi^{\phi})^{2}+\pi\pi^{\phi}] \nonumber  \\
& &
-\gamma^{1/2}e^{-\phi}R-\gamma^{1/2}e^{-\phi}\gamma_{ij}\partial_{i}
{\phi}\partial_{j}{\phi}+2\gamma^{1/2}\triangle e^{-\phi}=0 ~~~,
\end{eqnarray}

\noindent and to the momentum constraints

\begin{equation}\label{eq:5} 
{\mathcal{H}}_{i}=-2(\gamma_{ik}\pi^{kj})_{,j}+\pi^{kl}
\gamma_{kl,i}+\pi^{\phi}\phi_{,i}=0 ~~~.
\end{equation}

      Variation with respect to the canonical variables yield the evolution
equations

\begin{equation}\label{eq:6}
-2K_{ij}=\frac{1}{N} \left( \dot{\gamma}_{ij}-N_{i;j}-N_{j;i} \right) 
=\gamma^{-1/2}e^{\phi}
\left( 2\pi^{kl}\gamma_{ik}\gamma_{jl}+\gamma_{ij}\pi^{\phi} \right) ~~~, 
\end{equation}

\begin{equation}\label{eq:7}
\frac{1}{N} \left( \dot{\phi}-N^{i}\phi_{,i} \right) 
=\gamma^{-1/2}e^{\phi} \left( \pi^{\phi}+\pi \right) 
~~~,
\end{equation}

\begin{equation}\label{eq:8}
\frac{1}{N} \left[ \dot{\pi}^{i}_{\,j}- \left( N^{m}\pi^{i}_{\,j} \right) _{,m}
+N^{i}_{\, ,m}\pi^{m}_{\,j}
-N^{m}_{\,\, ,j}\pi^{i}_{\,m} \right] 
=\frac{1}{2}\gamma^{-1/2}e^{\phi}\delta^{i}_{\,j}
\left[ \pi^{mn}\pi^{kl}\gamma_{mk}\gamma_{nl}
+\frac{1}{2} \left(\pi^{\phi} \right) ^{2}
+\pi\pi^{\phi} \right] ~~~,
\end{equation} 

\begin{equation}\label{eq:9}
\frac{1}{N} \left[ \dot{\pi}^{\phi}-(N^{i}\phi^{\phi})_{,i} \right] 
=-\gamma^{-1/2}e^{\phi}
\left[ \pi^{ij}\pi^{kl}\gamma_{ik}\gamma_{jl}
+\frac{1}{2} \left(\pi^{\phi} \right) ^{2}
+\pi\pi^{\phi} \right] ~~~.
\end{equation}

\noindent Here $K_{ij}$ is the extrinsic curvature, which is the relevant 
object in lagrange formalism. 

     The basic idea of canonical transformations is well known
\cite{art7}. One defines new fields, which we denote by tilde, so that
Hamilton's equations are preserved. This implies that the new action has
the same form as the original except that it may have a total time
derivative added to it,

\begin{equation}\label{eq:10.1}
\Gamma=\int(\tilde{\pi}^{ij}\dot{\tilde{\gamma}}_{ij}+\tilde{\pi}^{\phi}
\dot{\tilde{\phi}}-N\tilde{\mathcal{H}}-N^{i}\tilde{\mathcal{H}}_{i})
d^{4}x+\int\dot{\mathcal{S}}dt ~~~,
\end{equation}

\noindent where $\mathcal{S}$ is a functional which depends on the old and
the new field variables. It is assumed that $\mathcal{S}$ does not depend
on time explicitly. Applying the chain rule,

\begin{equation}\label{eq:10.2}
\dot{\mathcal{S}}=\int\left[\frac{\delta \mathcal{S}}{\delta
\phi(x)}\dot{\phi}(t,x)    
+\frac{\delta \mathcal{S}}{\delta \tilde{\phi}(x)}\dot{\tilde{\phi}}(t,x)
+\frac{\delta \mathcal{S}}{\delta \gamma_{ij}(x)}\dot{\gamma_{ij}}(t,x)
+\frac{\delta \mathcal{S}}{\delta
\tilde{\gamma}_{ij}(x)}\dot{\tilde{\gamma}_{ij}}(t,x)
\right]d^{3}x ~~~,
\end{equation}

\noindent and comparing equation (\ref{eq:10.2}) with (\ref{eq:10.1}), one
derives the canonical transformation linking the various variables,

\begin{equation}\label{eq:10.3}
{\mathcal{H}}(x)=\tilde{\mathcal{H}}(x)  ~~~, \hspace{.2in}
{\mathcal{H}}_{i}(x)=\tilde{\mathcal{H}}_{i}(x)
\end{equation}

\begin{eqnarray}\label{eq:10.4}
\pi^{\phi}(x)=\frac{\delta \mathcal{S}}{\delta \phi(x)} ~~~~,~~~~
\pi^{ij}=\frac{\delta \mathcal{S}}{\delta \gamma_{ij}(x)} ~~~, \\
\tilde{\pi}^{\phi}(x)=-\frac{\delta \mathcal{S}}{\delta \tilde{\phi}(x)}
~~~~,~~~~
\tilde{\pi}_{ij}=-\frac{\delta \mathcal{S}}{\delta \tilde{\gamma}_{ij}(x)}
~~~. \nonumber
\end{eqnarray}

\noindent The new variables, denoted by a tilde, will be chosen 
so that the new Hamiltonian density vanishes strongly. Therefore 
they are constant in time.

\section{Hamilton-Jacobi Equation}

    The HJ equation is given by the hamiltonian constraint (\ref{eq:4}) 
after expressing the momenta through equation (\ref{eq:10.4}). In the
long-wavelength limit we neglect the last three terms since they involve
two spatial derivatives. We thus obtain the HJ equation in the 
approximate term

\begin{equation}\label{eq:11}
\gamma^{-1/2}e^{\phi} \left[ \frac{\delta \mathcal{S}}{\delta \gamma_{ij}}
\frac{\delta \mathcal{S}}{\delta \gamma_{kl}} \gamma_{ik} \gamma_{jl}
+\frac{1}{2}(\frac{\delta \mathcal{S}}{\delta \phi})^{2}
+\gamma_{ij}\frac{\delta \mathcal{S}}
{\delta \gamma_{ij}}\frac{\delta \mathcal{S}}{\delta \phi} \right] =0
~~~. 
\end{equation}

    The HJ equation is interesting because of its intimate
relation to quantum gravity. The Wheeler-De Witt equation, and the momentum 
constraint for the effective string action are given by

\begin{equation}\label{eq:11.1}
{\mathcal{H}} \Psi =0  ~~~~~~, \hspace{.6in} {\mathcal{H}}_{i} \Psi =0 ~~~,
\end{equation}

\noindent  where the canonical commutation relations,

\begin{eqnarray}\label{eq:11.2}
\left[ \gamma _{ij}(x) ,\pi^{kl}(x') \right] & = & \frac{i}{2}
(\delta^{k}_{\,i}
\delta^{l}_{\,j}+\delta^{k}_{\,j} \delta^{l}_{\,i} ) \delta (x-x') ~~~, \\
\left[ \phi(x),\pi^{\phi}(x') \right]  & = & i \delta (x-x') ~~~, \nonumber
\end{eqnarray}

\noindent are used. If we consider the WKB approximation,
we get the HJ equation at the lowest order.

    The HJ equation takes a simpler form by using, instead 
of $\phi$, the shifted dilaton $\Phi = \phi - \ln {\gamma^{1/2}}$. We 
then find

\begin{equation}\label{eq:11.3}
e^{\Phi} \left[ Tr \left( \gamma\frac{\delta \mathcal{S}}
{\delta \gamma} \gamma
\frac{\delta \mathcal{S}}{\delta \gamma} \right)
-\frac{1}{4}(\frac{\delta \mathcal{S}}{\delta \Phi})^{2} \right]=0
~~~. 
\end{equation}

\noindent Equation (\ref{eq:11.3}) is invariant under SFD transformation,
which is defined as,
  
\begin{equation}\label{eq:11.5.1}
\left[ \gamma \right] \longrightarrow \left[ \gamma \right]^{-1} ~~~, 
\hspace{.2in}
\Phi \longrightarrow \Phi ~~~. \nonumber
\end{equation}

\noindent Notice that, although the HJ equation is SFD invariant, 
the momentum constraints are not. We can write the 
momentum constraint, via the equations (\ref{eq:5}) and (\ref{eq:10.4})

\begin{equation}\label{eq:11.5.2}
{\mathcal{H}}_{i}=-2 \left( \gamma_{ik} \frac{\delta \mathcal{S}}
{\delta \gamma_{kj}} \right)_{,j}
+\frac{\delta \mathcal{S}}{\delta \gamma_{kl}}  \gamma_{kl,i}
+\frac{\delta \mathcal{S}}{\delta \phi}\phi_{,i}=0 ~~~. 
\end{equation}

\noindent They state that the generating functional is
invariant under spatial diffeomorphisms.

\section{General Solution near a Singularity}

\subsection{Ansatz for the Generating Functional}
   
     In this section, we investigate the full classical long-wavelength
problem of the low energy string cosmology. The generating functional
$\mathcal{S}$ is assumed to be a function of both the scalar field and the
gravitational field. Adapting an ansatz used by Salopek \cite{art7}, 
the dependence on the gravitational field is chosen in a specific way, 
without losing the generality of the solution.
   
    We choose the ansatz given below for the lowest order generating
functional, 
    
\begin{equation}\label{eq:12}
{\mathcal{S}}^{o}=-2\int
e^{-\frac{3}{2}(\phi-\tilde{\phi})}H(\phi,h_{ij}
;\tilde{\phi},\tilde{h}_{ij})
\sqrt{\gamma}d^{3}x ~~~,
\end{equation}

\noindent where $h_{ij}=\gamma^{-1/3}\gamma_{ij}$, and
$\tilde{h}_{ij}=\tilde{\gamma}^{-1/3}\tilde{\gamma}_{ij}$ are the
unimodular conformal three-metric.  Note that we introduced the
$e^{-\frac{3}{2}\tilde{\phi}}$ factor only for convenience, as it will
become clear in the following. Since

\begin{equation}\label{eq:12.1}
\frac{\partial H}{\partial \gamma_{ij}}=\gamma^{-1/3}
\left[\frac{\partial H}{\partial h_{ij}}-\frac{1}{3}
\frac{\partial H}{\partial h_{kl}}h_{kl}h^{ij}\right] ~~~,
\end{equation}

\noindent  We can write the HJ equation in terms of 
the conformal metric $h_{ij}$,

\begin{equation}\label{eq:13}
H^{2}=\frac{8}{3}\frac{\partial H}{\partial h_{ij}}\frac{\partial
H}{\partial h_{kl}} \left( h_{ik}h_{jl}-\frac{1}{3}h_{ij}h_{kl} \right)
+\frac{4}{3} \left( \frac{\partial H}{\partial \phi} \right) ^{2} ~~~.
\end{equation}

\noindent This is referred to as the separated HJ equation.
    
     Inserting (\ref{eq:12}) in (\ref{eq:10.4}) we find the new and the old 
momenta

\begin{equation}\label{eq:14} 
\pi^{ij}=-2\gamma^{1/2}e^{-\frac{3}{2}(\phi-\tilde{\phi})}
\left[ \frac{1}{2}\gamma^{ij}H+\gamma^{-1/3} \left( 
\frac{\partial H}{\partial h_{ij}}
-\frac{1}{3}\frac{\partial H}{\partial h_{kl}}h_{kl}h_{ij} \right) \right] ~~~,
\end{equation}

\begin{equation}\label{eq:15}
\tilde{\pi}^{ij}=
2\gamma^{1/2}\tilde{\gamma}^{-1/3}e^{-\frac{3}{2}(\phi-\tilde{\phi})}
\left[ \frac{\partial H}{\partial \tilde{h}_{ij}}-\frac{1}{3}\frac{\partial H}
{\partial \tilde{h}_{kl}}\tilde{h}_{kl}\tilde{h}_{ij} \right] ~~~,
\end{equation}

\begin{equation}\label{eq:16}
\pi^{\phi}=-2 \gamma^{1/2} 
e^{-\frac{3}{2}
(\phi-\tilde{\phi})} \left[ -\frac{3}{2}H+\frac{\partial H}{\partial \phi}
\right] ~~~,
\end{equation}

\begin{equation}\label{eq:17}
\tilde{\pi}^{\phi}=2\gamma^{1/2}e^{-\frac{3}{2}(\phi-\tilde{\phi})}
\left[ \frac{3}{2}H
+\frac{\partial H}{\partial \tilde{\phi}} \right] ~~~.
\end{equation}

\noindent The trace of the gravitational momentum is proportional to H,

\begin{equation}\label{eq:18}
\pi=\pi^{i}_{\,i}=-3\gamma^{1/2}e^{-\frac{3}{2}(\phi-\tilde{\phi})}H ~~~.
\end{equation}

\noindent and this is proportional to the integrand in equation (\ref{eq:12}).

     The specific choice for the dependence of the function H on the
metric, through the combination $h_{ij}=\gamma^{-1/3}\gamma_{ij}$,
effectively decomposes the gravitational momentum tensor into a trace
contribution and a traceless part which describes the evolution of
gravitational radiation. Similarly, the new gravitational momentum tensor
is traceless, $\tilde{\pi}^{ij}\tilde{\gamma}_{ij}=0$.

      Furthermore, we attempt the following solution to the separated
HJ equation (\ref{eq:13})

\begin{equation}\label{eq:19}
H(\phi, h_{ij};\tilde{\phi},\tilde{h}_{ij})\equiv 
H(\phi,\tilde{\phi},z) ~~~,
\end{equation}

\noindent where $z$ is defined as \cite{art7},
 
\begin{equation}\label{eq:20}
z^{2}=\frac{1}{2}Tr \left[ \ln \left( [h][\tilde{h}]^{-1} \right)
\ln \left([h][\tilde{h}]^{-1} \right) \right] ~~~. \nonumber
\end{equation}

\noindent Here $[h]$ and $[\tilde{h}]^{-1}$ are matrices with components
$h_{ij}$ and $\tilde{h}^{ij}$ respectively. This variable may be thought
of as the ``distance'', in field space, between the old conformal metric
$h_{ij}$ and the new one $\tilde{h}_{ij}$, \cite{art7}, \cite{art10}. We
do not lose any information, because six constants of integration have
been introduced through $\tilde{h}_{ij}$, which are sufficient to describe
the dynamics of the gravitational field. 

    In terms of this variable, the separated Hamilton-Jacobi 
equation reduces to that of massless scalar fields $z$ and $\phi$,

\begin{equation}\label{eq:21}
H^{2}=\frac{4}{3} \left[ (\frac{\partial H}{\partial
z})^{2}+(\frac{\partial
H}{\partial \phi})^{2} \right] ~~~,
\end{equation}

\noindent where, we used

\begin{equation}\label{eq:22}
\frac{\partial H}{\partial h_{ij}}=\frac{\partial H}{\partial z}
\frac{\partial z}{\partial h_{ij}}=\frac{1}{2}\frac{\partial H}{\partial z}
z^{-1} \left[ [h]^{-1}\ln \left( [h][\tilde{h}]^{-1} \right) \right] ^{ij} ~~~.
\end{equation}

\noindent Similarly

\begin{equation}\label{eq:23}
\frac{\partial H}{\partial \tilde{h}_{ij}}=\frac{\partial H}{\partial z}
\frac{\partial z}{\partial \tilde{h}_{ij}}=-\frac{1}{2}\frac{\partial H}
{\partial z}
z^{-1} \left[ [h]^{-1}\ln \left( [h][\tilde{h}]^{-1} \right) [h]
[\tilde{h}]^{-1} \right] ^{ij} ~~.
\end{equation}

\noindent Equations (\ref{eq:22}) and (\ref{eq:23}) yield

\begin{eqnarray}\label{eq:24}
h_{ij}\frac{\partial H}{\partial h_{ij}}=0 ~~~~, \hspace{.5in}
\tilde{h}_{ij}\frac{\partial H}{\partial \tilde{h}_{ij}}=0 ~~~~.
\end{eqnarray}

\noindent Therefore the equations for the momenta become

\begin{equation}\label{eq:25} 
\pi^{ij}=-2\gamma^{1/2}e^{-\frac{3}{2}(\phi-\tilde{\phi})}
\left[ \frac{1}{2}\gamma^{ij}H+\gamma^{-1/3}
\frac{\partial H}{\partial h_{ij}} \right] ~~~,
\end{equation}

\begin{equation}\label{eq:26}
\tilde{\pi}^{ij}=
2\gamma^{1/2}\tilde{\gamma}^{-1/3}e^{-\frac{3}{2}(\phi-\tilde{\phi})}
\frac{\partial H}{\partial \tilde{h}_{ij}} ~~~,
\end{equation}

\begin{equation}\label{eq:27}
\pi^{\phi}=-2 \gamma^{1/2} 
e^{-\frac{3}{2}
(\phi-\tilde{\phi})} \left[ -\frac{3}{2}H
+\frac{\partial H}{\partial \phi} \right] ~~~,
\end{equation}

\begin{equation}\label{eq:28}
\tilde{\pi}^{\phi}=2\gamma^{1/2}e^{-\frac{3}{2}(\phi-\tilde{\phi})}
\left[ \frac{3}{2}H
+\frac{\partial H}{\partial \tilde{\phi}} \right] ~~~.
\end{equation}

\noindent Note that the new gravitational momentum is related to the old
one through the reciprocity relation

\begin{equation}\label{eq:28.1}
\pi^{ij}\gamma_{jl}=\frac{1}{3}\pi\delta^{i}_{\,l}
+\tilde{\pi}^{ij}\tilde{\gamma}_{jl} ~~~.
\end{equation}

     We will also need the equations for $z$ and its momentum. One can 
write them in a manner similar to those of the dilaton. 

\begin{equation}\label{eq:29}
\dot{z}=\gamma^{-1/2}e^{\phi}(\pi^{z}+\pi) ~~~,
\end{equation}

\begin{equation}\label{eq:30}
\pi^{z}=-2\gamma^{1/2}e^{-\frac{3}{2}(\phi-\tilde{\phi})} \left[ 
-\frac{3}{2}H
+\frac{\partial H}{\partial z} \right] ~~~.
\end{equation}

\noindent It will become clear, below, that these equations are consistent 
and provide the correct evolution for $z$.

\subsection{Momentum Constraint} 
  
    The momentum constraints admit a simple expression throught the 
solution (\ref{eq:19}), \cite{art7}. The gravitational momentum tensor 
can be decomposed into a trace and a traceless part which we denote by an 
overbar

\begin{equation}\label{eq:28.2}
\pi^{ij}=\frac{1}{3}\pi\gamma^{ij}+\bar{\pi}^{ij}~~~.
\end{equation}

\noindent The momentum constraints become

\begin{equation}\label{eq:28.3}
{\mathcal{H}}_{i}=-\frac{2}{3}\pi_{,i}-2(\bar{\pi}^{jl}\gamma_{li})_{,j}
+\pi^{kl}\gamma_{kl,i}+\pi^{\phi}\phi_{,i}=0 ~~~.
\end{equation}

\noindent Using equation (\ref{eq:18}), the generating functional 
can be written in terms of the trace of the gravitational momentum

\begin{equation}\label{eq:28.4}
{\mathcal{S}}=\frac{2}{3}\int\pi\left(\phi(x),h_{ij}(x)
;\tilde{\phi}(x),\tilde{h}_{ij}(x)\right)d^{3}x ~~~.
\end{equation}

\noindent Therefore the new and the old canonical variables can be 
expressed as partial derivatives of $\pi$. The spatial derivative of $\pi$
can be written as

\begin{equation}\label{eq:28.5}
\pi_{,i}=\frac{3}{2}\pi^{kl}\gamma_{kl,i}
-\frac{3}{2}\tilde{\pi}^{kl}\tilde{\gamma}_{kl,i}
+\frac{3}{2}\pi^{\phi}\phi_{,i}
-\frac{3}{2}\tilde{\pi}^{\phi}\tilde{\phi}_{,i}~~~.
\end{equation}

\noindent If we substitute this in to the equation (\ref{eq:28.3}), using the
reciprocity relation, we can write the momentum constraint in terms of the 
new variables

\begin{equation}\label{eq:28.6} 
\tilde{\mathcal{H}}_{i}=-2(\tilde{\gamma}_{ik}\tilde{\pi}^{kj})_{,j}
+\tilde{\pi}^{kl}
\tilde{\gamma}_{kl,i}+\tilde{\pi}^{\phi}\tilde{\phi}_{,i}=0 ~~~.
\end{equation}

\noindent Here one effectively performs a legendre transformation between
the new and the old variables.

    The evolution equations for the new variables are given by the new action  
which was written in equation (\ref{eq:10.1})

\begin{equation}\label{eq:28.7}
\tilde{\Gamma}=\int(\tilde{\pi}^{ij}\dot{\tilde{\gamma}}_{ij}
+\tilde{\pi}^{\phi}
\dot{\tilde{\phi}}-N^{i}\tilde{H}_{i})d^{4}x ~~~. 
\end{equation}

\noindent One can easily see that, if the shift function $N_{i}$ vanishes,
then the new canonical variables are independent of time, but they can
depend on space coordinates. They are restricted by the momentum
constraint (\ref{eq:28.6}). We can write the momentum constraint in terms
of $\tilde{h}_{ij}$,
   
\begin{equation}\label{eq:28.8}
\tilde{\mathcal{H}}_{i}=-2(\tilde{\gamma}^{1/3}
\tilde{h}_{ik}\tilde{\pi}^{kj})_{,j}
+\tilde{\gamma}^{1/3}\tilde{\pi}^{kl}
\tilde{h}_{kl,i}+\tilde{\pi}^{\phi}\tilde{\phi}_{,i}=0 ~~~.
\end{equation}
 
\noindent Since the theory does not depend on the parametrization of the
spatial coordinates, one may write the momentum constraint in terms of a
covariant derivative with respect to $\tilde{h}_{ij}$,

\begin{equation}\label{eq:28.9}
\tilde{\mathcal{H}}_{i}=-2(\tilde{\gamma}^{1/3}\tilde{\pi}^{j}_{i})_{;j}
+\tilde{\pi}^{\phi}\tilde{\phi}_{,i}=0 ~~~.
\end{equation}

\subsection{Solution}

    Solution of the equation (\ref{eq:21}) is given by

\begin{equation}\label{eq:31}
H=-\frac{2}{3t_{o} e^{\tilde{\phi}}} \exp \left\{ \frac{\sqrt{3}}{2}
\left[ \left( \phi-\tilde{\phi} \right) ^{2}+ \left( z-\tilde{z} \right) 
^{2} \right] ^{1/2} \right\} ~~~,
\end{equation}

\noindent where a tilde refers to initial value of the corresponding
variable. The initial value of $H$ is chosen in order to have a Pre-Big Bang
behaviour and $t_{o}$ is an arbitrary constant \cite{art9}. Its meaning will 
become appearent below. Here one should note the rotational symmetry of the 
solution.

    Using the evolution equations for $\phi$ and $z$ (\ref{eq:7}), 
(\ref{eq:29}), and the equations for their conjugate momentum 
(\ref{eq:27}), (\ref{eq:30}), we find

\begin{equation}\label{eq:33}
\dot{\phi}=-2e^{\phi}e^{-\frac{3}{2}(\phi-\tilde{\phi})}\frac{\partial
H}{\partial \phi} ~~~,
\end{equation}

\begin{equation}\label{eq:35}
\dot{z}=-2e^{\phi}e^{-\frac{3}{2}(\phi-\tilde{\phi})}\frac{\partial
H}{\partial z} ~~~.
\end{equation}

\noindent If we use the new variables $x$ and $y$ defined as

\begin{equation}\label{eq:36}
x=\frac{\sqrt{3}}{2}(\phi-\tilde{\phi}) ~~~, \hspace{.1in}
y=\frac{\sqrt{3}}{2}(z-\tilde{z}) ~~~, \hspace{.1in} 
r^{2}=x^{2}+y^{2} ~~~, \nonumber
\end{equation}

\noindent then we find  

\begin{eqnarray}\label{eq:38}
\dot{x}=\frac{1}{t_{o}}\frac{x}{\sqrt{x^{2}+y^{2}}}e^{-\frac{1}{\sqrt{3}}x
+\sqrt{x^{2}+y^{2}}} ~~~,  \\
\vspace{2.0in}
\dot{y}=\frac{1}{t_{o}}\frac{y}{\sqrt{x^{2}+y^{2}}}e^{-\frac{1}{\sqrt{3}}x
+\sqrt{x^{2}+y^{2}}} ~~~.
\end{eqnarray}

\noindent Because of rotational symmetry, it is natural to use the polar
coordinates in the $(\phi,z)$ plane. Then one finds that the angular
coordinate is constant in time.  It depends only on the spatial
coordinates. The radial coordinate is given by

\begin{equation}\label{eq:39.1}
r=\frac{-1}{1-\frac{1}{\sqrt{3}}\cos{\varphi}}
\ln \left( 1-\frac{t}{\tilde{t}} \right) ~~~, \hspace{.2in}
\tilde{t}=\frac{t_{o}}{1-\frac{1}{\sqrt{3}}\cos{\varphi}} ~~~.
\end{equation}

\noindent We find, using $x=r\cos{\varphi}$, $y=r\sin{\varphi}$ and
equation (\ref{eq:36}),

\begin{equation}\label{eq:43}
\phi=\tilde{\phi}+\beta\ln \left( 1-\frac{t}{\tilde{t}} \right) ~~~, 
\hspace{.1in}
\beta=\frac{-\frac{2}{\sqrt{3}} \cos{\varphi}}
{1-\frac{1}{\sqrt{3}} \cos{\varphi}} ~~~, 
\end{equation}

\vspace{.3in}

\begin{equation}\label{eq:45}
z=\frac{-\frac{2}{\sqrt{3}} \sin{\varphi}}
{1-\frac{1}{\sqrt{3}} \cos{\varphi}}
\ln \left( 1-\frac{t}{\tilde{t}} \right) ~~~.
\end{equation}

\noindent Here notice that $\tilde{z}=0$ by definition. In the
$(\varphi,z)-$plane circles concentric with the origin corresponds to
constant $H$ surfaces. The evolution of the fields $\phi$ and $z$ at a
fixed spatial point are given by the rays originating from the origin.
They remain orthogonal to the uniform $H$ surfaces everytime.  One can
see, using equations (\ref{eq:25}) and (\ref{eq:27}), that the momenta for
the gravitational field and the dilaton are constant in time.  But they
can have spatial dependence,
    
\begin{equation}\label{eq:32}
\pi^{i}_{\,j}=\lambda^{i}_{\,j}(x) ~~~.
\end{equation}

     The evolution of the unimodular conformal metric $h_{ij}$ 
is given by

\begin{equation}\label{eq:34}
\dot{h}_{ij}=-4e^{\phi}e^{-\frac{3}{2}(\phi-\tilde{\phi})}\frac{\partial H}
{\partial h_{kl}}h_{ki}h_{lj} ~~~,
\end{equation}

\noindent Here equations (\ref{eq:6}), (\ref{eq:25}) and (\ref{eq:27}) are
used. At this point, by a direct application of the chain rule, one can
check that the equations (\ref{eq:29}) and (\ref{eq:30}) for $\dot{z}$ and
$\pi^{z}$ are consistent such that they lead to correct expression for the
evolution of $z$.  We can find the evolution of the unimodular conformal
metric using equation (\ref{eq:34}) and the solution (\ref{eq:45}) for
$z$,

\begin{equation}\label{eq:45.1}
\left[ \ln \left( [h][\tilde{h}]^{-1} \right) \right] =z\left[ p(x) \right] 
~~~.
\end{equation}

\noindent Here the matrix $\left[ p(x) \right]$ satisfies

\begin{eqnarray}\label{eq:45.2}
Tr \left( [p][p] \right) =2 ~~~~, \hspace{.5in} Tr \left( [p] \right) =0 ~~~.
\end{eqnarray}

\noindent Hereafter, in order to avoid repetition, we are going to write 
the results without going into the details of algebraic manipulations.  
Using equation (\ref{eq:25}), we find

\begin{equation}\label{eq:54.1.1}
\frac{1}{\lambda} \left[ \lambda \right] 
=\frac{1}{3} \left( I+\frac{\sqrt{3}}{2}
\sin{\varphi}[p] \right)
\end{equation}

\noindent At this stage, eliminating
$[p]$ in favor of $[\lambda]$, we define the matrice $[\alpha]$,

\begin{equation}\label{eq:52}
[\alpha(x)]=I-\frac{2}{1-\frac{1}{\sqrt{3}}
\cos {\varphi}}\frac{1}{\lambda}[\lambda] ~~~.
\end{equation}

\noindent Then the trace of  $[\alpha]$ is

\begin{equation}\label{eq:53}
\alpha=\frac{1-\sqrt{3} \cos {\varphi}}
{1-\frac{1}{\sqrt{3}} \cos {\varphi}} ~~~.
\end{equation}

\noindent The matrice  $[\alpha]$ can be simplified further as below. Equation 
(\ref{eq:45.1}) yields

\begin{equation}\label{eq:46}
\left[ h(t,x) \right] =\exp \left\{ 2 \left( [\alpha]
-\frac{1}{3}\alpha I \right)
\ln \left( 1-\frac{t}{\tilde{t}} \right)
\right\}
\left[ \tilde{h}(x) \right] ~~~.
\end{equation}

\noindent If we replace equations (\ref{eq:25}) and (\ref{eq:27})  in
equation (\ref{eq:6}), then we find the determinant of the metric evolves
according to

\begin{equation}\label{eq:49}
\gamma= \left( \frac{\lambda \tilde{t} e^{\tilde{\phi}}}{3-\alpha} \right) ^{2}
\left( 1-\frac{t}{\tilde{t}} \right)^{2 \alpha} ~~~.
\end{equation}

\noindent Then we find the evolution of the metric
$\gamma_{ij}=\gamma^{1/3}h_{ij}$,

\begin{equation}\label{eq:50}
\left[ \gamma(t,x) \right] =\exp \left\{
2[\alpha(x)]\ln \left(1-\frac{t}{\tilde{t}} 
\right) \right\} 
\left[ \tilde{\gamma}(x) \right] ~~~,
\end{equation}

\noindent where

\begin{equation}\label{eq:51}
\left[ \tilde{\gamma}(x) \right]= \left( \frac{\lambda
\tilde{t} e^{\tilde{\phi}}}{3-\alpha} \right) ^{2/3} 
\left[ \tilde{h}(x) \right] ~~~.
\end{equation}

\noindent It is possible to introduce local coordinates in which the
matrice $[\alpha]$ and the metric are diagonal for discussion of duality
below. One can also write, arranging equations (\ref{eq:43}) and
(\ref{eq:45}), the scalar field $\phi$, and $z$ in terms of trace
$\alpha$. Here, it is illuminating to find the extrinsic curvature.  We
obtain, using equation (\ref{eq:6}),
 
\begin{equation}\label{eq:51.3}
\left[ K \right] =\frac{\left[ \alpha \right] }{\tilde{t}-t} ~~~,
\end{equation}

\noindent where $\left[ K \right]$ is the matrice representation of the 
extrinsic curvature, with entries $K^{i}_{\, j}$. This yields the 
expansion rate of the universe as
 
\begin{equation}\label{eq:51.4}
K=-\frac{1}{2}\frac{\dot{\gamma}}{\gamma}=\frac{\alpha}{\tilde{t}-t} ~~~.
\end{equation}

\noindent Meanwhile we can write H as, 

\begin{equation}\label{eq:48}
H=-\frac{3-\alpha}{3\tilde{t}e^{\tilde{\phi}}}
\left( 1-\frac{t}{\tilde{t}} \right)
^{-\frac{3-\alpha}{2}} ~~~.
\end{equation}
 
\noindent $K$ is proportional to $e^{-\frac{\phi}{2}}H$. Equation
(\ref{eq:53}) yields $-\sqrt{3}\leq\alpha\leq\sqrt{3}$. The condition for
quasi-homogeneous regions to undergo superinflation, $\alpha <0$,
corresponds to the region $\cos{\varphi}>1/\sqrt{3}$ in the
$(\phi,z)-$plane \cite{art9}. The maximal rate of expansion is reached for
$\alpha=-\sqrt{3}$.  This corresponds to the quasi-isotropic case
$\varphi=0$, as explained in the next chapter.  $[\alpha]$ and the matrix
of gravitational momentum $[\lambda]$ are related as

\begin{equation}\label{eq:54}
\frac{1}{\lambda}[\lambda]=\frac{1}{3-\alpha} \left(I-[\alpha] \right) ~~~.
\end{equation}

\noindent The relations (\ref{eq:45.2}) and  (\ref{eq:54.1.1}) yields

\begin{equation}\label{eq:54.1}
\frac{1}{\lambda^{2}}\lambda^{i}_{\,j}\lambda^{j}_{\,i}
=\frac{1}{6} \left( 3-\cos^{2}{\varphi} \right) ~~~,
\end{equation}

\noindent and this gives the condition 

\begin{equation}\label{eq:54.2}
\frac{1}{3}\leq \frac{1}{\lambda^{2}}
\lambda^{i}_{\,j}\lambda^{j}_{\,i}\leq \frac{1}{2} ~~~.
\end{equation}

\noindent Here, the lower limit corresponds to the quasi-isotropic 
case. Therefore we have a Kasner-like solution,

\begin{eqnarray}\label{eq:55}
Tr \left( [\alpha][\alpha] \right) =1 ~~~,\hspace{.5in} \beta=-1+\alpha ~~~.
\end{eqnarray}

    Momenta $\pi^{\phi}$ and $\pi^{z}$ are related to angle $\varphi$ and 
trace of the gravitational momentum $\lambda$ as follows

\begin{eqnarray}\label{eq:56}
\pi^{\phi}= \left( -1+\frac{1}{\sqrt{3}} \cos {\varphi} \right) \lambda
~~~,\hspace{.2in}
\pi^{z}= \left( -1+\frac{1}{\sqrt{3}} \sin {\varphi} \right) \lambda ~~~,
\end{eqnarray}

\noindent and they satisfy

\begin{equation}\label{eq:57}
\left( \pi^{\phi}+\lambda \right) ^{2}+ \left( \pi^{z}+\lambda \right) 
^{2}=\frac{1}{3}\lambda^{2} ~~~.
\end{equation}

\noindent Initial momentum for $z$, $\phi$, and the gravitational field are

\begin{eqnarray}\label{eq:58}
\tilde{\pi}^{\phi}=\pi^{\phi} ~~~~, \hspace{.5in} \tilde{\pi}^{z}=\pi^{z} ~~~,
\end{eqnarray}

\begin{equation}\label{eq:59}
\tilde{\pi}^{i}_{\,j}=\lambda^{i}_{\,j}-\frac{1}{3}\delta^{i}_{\,j}
\lambda ~~~.
\end{equation}

\section{Quasi-isotropic Solution}

     In this section we consider a quasi-isotropic space. This is a
special case of the general solution, as it is explained below.  If we use
the quasi-isotropic ansatz

\begin{equation}\label{eq:11.6}
{\mathcal{S}}^{o}=-2\int
e^{-\frac{3}{2}\phi}H(\phi)
\sqrt{\gamma}d^{3}x ~~~,
\end{equation}

\noindent the Hamilton-Jacobi equation reduces to

\begin{equation}\label{eq:11.7}
H^{2}=\frac{4}{3}(\frac{\partial H}{\partial \phi})^{2} ~~~.
\end{equation}

\noindent The momentum constraint (diffeomorphism invariance)

\begin{equation}\label{eq:11.8}
H_{,i}=\frac{\partial H}{\partial \phi}\phi_{,i} ~~~,
\end{equation}

\noindent is automatically satisfied by this ansatz.
Meanwhile the equations of motion for the fields are

\begin{equation}\label{eq:11.9}
\dot{\phi}=-2e^{-\frac{1}{2} \phi}\frac{\partial H}{\partial \phi} ~~~, 
\hspace{.4in}
\dot{\gamma}_{ij}=e^{-\frac{1}{2}
\phi}\left[H-2\frac{\partial H}
{\partial \phi} \right] \gamma_{ij} ~~~.
\end{equation}

\noindent They immediately yield a quasi-isotropic solution 
$\gamma_{ij}=a^{2}(\phi)h_{ij}(x)$. Here 

\begin{equation}\label{eq:11.10}
a^{2}=\exp \left\{ -\frac{1-\sqrt{3}}{2}\int 
\frac{H}{\partial H / \partial \phi} d\phi \right\} ~~~,
\end{equation}

\noindent and $h_{ij}$ is the seed metric. If we solve the equations and the 
constraints explicitly, we find, for the scalar field and the gravitational
field

\begin{eqnarray}\label{eq:11.11}
\phi &=& \frac{2}{1-\sqrt{3}}
\ln \left( 1-\frac{t}{\tilde{t}} \right) ~~~, \\ \\
\left[ \gamma \right] &=& \left( 1-\frac{t}{\tilde{t}} \right)
^{-\frac{2}{\sqrt{3}}} \left[ h(x) \right] ~~~. \nonumber
\end{eqnarray}

\noindent Here $t_{o}$ is rescaled by a factor of $(1-1/\sqrt{3})^{-1}$. 
We obtain, using equation (\ref{eq:10.4}) and the solution, that 
the momentum of the gravitational field and the scalar field are constant 
independent of space-time coordinates. Furthermore the traceless 
part of the gravitational momentum is zero and evolution of the 
gravitational radiation is neglected. $H$ is given by

\begin{equation}\label{eq:11.12}
H=\frac{2}{\sqrt{3}(1-\sqrt{3})}\frac{1}{\tilde{t}} 
\left( 1-\frac{t}{\tilde{t}} \right)
^{\frac{\sqrt{3}}{1-\sqrt{3}}} ~~~.
\end{equation}

\noindent Meanwhile we obtain for the extrinsic curvature,

\begin{equation}\label{eq:11.121}
\left[ K \right] =-\frac{1}{\sqrt{3}}\frac{1}{\tilde{t}-t}I ~~~.
\end{equation}

\noindent This yields, for the expansion rate of the quasi-isotropic 
universe,

\begin{equation}\label{eq:11.122}
K=-\frac{\sqrt{3}}{\tilde{t}-t} ~~~.
\end{equation}

    We expect that the general solution contains the quasi-isotropic one
as a special case.  If, in the general solution, we consider the case
$\varphi=0$, then traceless part of $[\alpha]$ and the momentum
$[\lambda]$ disappear. They contain only trace parts which are independent
of space coordinates. The unimodular conformal metric $[h]$ becomes
independent of time. One should notice that, we introduced the factor
$e^{-\frac{3}{2}{\tilde{\phi}}}$ in the generating function by hand, only
for convenience. As a result the metric $[\gamma]$ and the scalar field
$\phi$ reduce to those of the quasi-isotropic space. Similarly, the
extrinsic curvature reduces to the corresponding quasi-isotropic one. This
correspondance can be checked explicitly, by putting $\phi=0$ in the
general solution. 

\section{Duality}

  In this section we briefly discuss the duality property of the general
solution. It is apparent, from the form of the solution, that the
transformation $\alpha_{a}\longrightarrow \alpha '_{a} =-\alpha_{a}$ (for
all $a$) generates a dual solution. This yields $\alpha\longrightarrow
\alpha=-\alpha$ for the trace of the matrice $[\alpha]$. We can perform
this as a transformation of the angular coordinate $\varphi$ in the
$(\phi,z)-$plane

\begin{equation}\label{eq:61}
\cos{(\pi-\varphi')}=\frac{\cos{\varphi}-\frac{\sqrt{3}}{2}}
{1-\frac{\sqrt{3}}{2}\cos{\varphi}} ~~~.
\end{equation}

\noindent One can easily check that, this transformation is equivalent to
$\alpha \longrightarrow \alpha '$, using equation (\ref{eq:53})
explicitly.  The dual solution is of the same form. This transformation is
well defined since $-1<\cos{\varphi'}<1$, and
$-\sqrt{3}<\alpha'<\sqrt{3}$. Transformation of the radius can be found by
using equation (\ref{eq:39.1}). Using $x=r\cos{\varphi}$ and
$y=r\sin{\varphi}$ we find the known result

\begin{equation}\label{eq:65}
\phi\longrightarrow \phi'=\phi-\ln{\gamma}
\end{equation}

\noindent However, $z$ does not experience any change except
an additional constant contribution (remember $\tilde{z}=0$ by definition). 
This can be seen easier if it is written in terms of $\alpha$.

     We can decompose the gravitational momentum tensor into a trace 
contribution and a traceless part as

\begin{equation}\label{eq:66}
\frac{1}{\lambda}[\lambda]=\frac{1}{3}I+[q] ~~~.
\end{equation}

\noindent Then we find that the traceless part $[q]$ transforms as

\begin{equation}\label{equ:67}
[q']=\frac{-\frac{1}{2}-\frac{\sqrt{3}}{2}\sqrt{1-6|q|^{2}}}
{-1-\frac{\sqrt{3}}{2}\sqrt{1-6|q|^{2}}}[q] ~~~,
\end{equation}

\noindent under the duality transformation. Here $|q|^{2}=Tr([q][q])$.
We should also impose the momentum constraints simultaneously. 

   Meanwhile, equation (\ref{eq:61}) yields that the dual of the 
quasi-isotropic solution is again a quasi-isotropic one. However it is 
not contained in the superinflationary section of the plane.

\section {Conclusion and Summary}

     In this paper, we applied the long-wavelength approximation to
low-energy effective string action in the context of Hamilton-Jacobi theory. 
The Hamilton-Jacobi equation for the effective string action in four dimensions
is invariant under SFD transformation. However the momentum constraints are not
invariant under this transformation. Long-wavelength gravity
(gradient expansion) has proved to be a fruitfull method for studying 
slightly inhomogeneous cosmology.It is a significant improvement 
over those of homogeneous minisuperspace.

     We presented leading order solution of HJ equation. This is the most
general solution near a singularity. We solved the HJ
equation including the evolution of the gravitational radiation. In order
to do this we performed a transformation to new canonical variables where
the Hamiltonian density vanishes strongly. Therefore, the new variables
are constant in time, if the shift function vanishes. However, they depend
on the spatial coordinates. In the separated Hamilton-Jacobi equation, the
gravitational degrees of freedom can be reduced to that of a single
massless scalar field. However, the gravitational field is fundamentally
different from massless scalar fields. For example, it carries spin
angular momentum, and the momentum constraints restrict the longitudinal
modes of the gravitational momentum tensor.
Then we presented the quasi-isotropic solution. The general solution 
contains this one as a special case. In this
case, gravitational radiation is neglected. 
The Hamilton-Jacobi approach yields a result consistent with the 
one derived by using the Lagrange equations directly, \cite{art9}.

   In the Hamilton-Jacobi approach, we can simply represent physically 
important cases in the $(\phi,z)-$plane. Constant $H$ surfaces are circles
concentric with the origin. The evolution of the fields $\phi$ and $z$, at
a fixed spatial point are given by trajectories originating from the
origin. These trajectories remain orthogonal to uniform $H$ surfaces 
everytime. The region $\cos\varphi>1/\sqrt{3}$ corresponds to 
superinflationary solutions. Meanwhile, we obtain a quasi-isotropic universe 
on the $\phi-$axis.  The Big Bang instant corresponds to a point at infinity. 
The momentum constraints admit a simple expression in terms of the new
canonical variables. This form is useful for general discussions. We also
performed the duality transformation as a transformation of angle in
$(\phi,z)-$plane, $\varphi \longrightarrow \varphi '$. We have to impose
the momentum constraints simultaneously. However this transformation and
its relation to the momentum constraint needs further clarification. 

\begin{center}
\textbf{Acknowledgments}
\end{center}
   
   I would like to thank CERN Theory Division and especially G. Veneziano
for proposing this problem, and for helpful discussions. I also thank
T\"{U}B\.{I}TAK (Scientific and Technical Research Council of Turkey) for
providing financial support during this study.

\vspace*{.5cm}

\newpage

\end{document}